\documentclass[unnumsec,webpdf,contemporary,large]{oup-authoring-template}
\pdfoutput=1
\graphicspath{{Fig/}}
\theoremstyle{thmstyleone}

\theoremstyle{thmstyletwo}%
\theoremstyle{thmstylethree}%

\usepackage{hyperref}
\usepackage{amssymb}
\usepackage{tabularx}
\usepackage{colortbl}
\usepackage{soul} 
\usepackage{color, xcolor} 

\colorlet{lightred}{red!30}

\sethlcolor{lightred} 
\soulregister{\cite}7 
\soulregister{\citep}7 
\soulregister{\citet}7 
\soulregister{\ref}7 
\soulregister{\pageref}7 

\begin{document}

\journaltitle{Briefings in Bioinformatics}
\DOI{DOI HERE}
\copyrightyear{2024}
\pubyear{2019}
\access{Advance Access Publication Date: Day Month Year}
\appnotes{Paper}

\firstpage{1}

\title[HiPM]{Adapting Differential Molecular Representation with Hierarchical Prompts for Multi-label Property Prediction}

\author[]{Linjia Kang\ORCID{0009-0008-2971-7087}${\text{\dag}}$}
\author[]{Songhua Zhou\ORCID{0009-0008-0276-2590}${\text{\dag}}$}
\author[]{Shuyan Fang\ORCID{0009-0003-1462-5124}}
\author[$\ast$]{Shichao Liu\ORCID{0000-0001-7217-4462}}

\authormark{Kang et al.}

\corresp[$\ast$]{Corresponding authors: 
Shichao Liu, E-mail: scliu@mail.hzau.edu.cn.\\
College of Informatics, Huazhong Agricultural University, Wuhan, Hubei 430070, China.\\
${\text{\dag}}$The authors wish it to be known that, in their opinion, the first two authors should be regarded as Joint First Authors.}

\received{Date}{0}{Year}
\revised{Date}{0}{Year}
\accepted{Date}{0}{Year}

\abstract{Accurate prediction of molecular properties is crucial in drug discovery. Traditional methods often overlook that real-world molecules typically exhibit multiple property labels with complex correlations. To this end, we propose a novel framework, HiPM, which stands for \textbf{Hi}erarchical \textbf{P}rompted \textbf{M}olecular representation learning framework. HiPM leverages task-aware prompts to enhance the differential expression of tasks in molecular representations and mitigate negative transfer caused by conflicts in individual task information. Our framework comprises two core components: the Molecular Representation Encoder (MRE) and the Task-Aware Prompter (TAP). MRE employs a hierarchical message-passing network architecture to capture molecular features at both the atom and motif levels. Meanwhile, TAP utilizes agglomerative hierarchical clustering algorithm to construct a prompt tree that reflects task affinity and distinctiveness, enabling the model to consider multi-granular correlation information among tasks, thereby effectively handling the complexity of multi-label property prediction. Extensive experiments demonstrate that HiPM achieves state-of-the-art performance across various multi-label datasets, offering a novel perspective on multi-label molecular representation learning.
}



\maketitle

\section{Introduction}
Traditional drug discovery requires an average of 10 to 15 years and costs over 2 billion dollars \cite{berdigaliyev2020overview, Catacutan2024}. However, recent advancements in molecular representation learning have revolutionized this field, significantly reducing both the time and cost \cite{chen2018rise, deng2022artificial, vamathevan2019applications}. Previous studies \cite{MolCLR, 10.1093/bioinformatics/btae118, AEGNN_M, 2024Zhang} typically represent molecules as topological graphs and employ Graph Neural Networks (GNNs) to capture both molecular structure and chemical information. This strategy has been widely validated as effective for learning molecular representations. Nonetheless, real-world molecules usually exhibit multiple properties and existing research often overlook the specific problems in this scenario. This limitation can result in significant gaps in understanding the complete biological activity of molecules, ultimately hindering the efficiency of drug discovery.

Considering molecular property prediction as a multi-label learning task is promising in resolving the above challenge. However, several inherent issues in multi-label learning must be addressed. \pagebreak One primary issue is the exponential growth of the output space. For instance, 32 labels can lead to as many as $2^{32}$ combinations. The other issue is gradient conflict \cite{gradient_surgery, LiuCVPR2023}. Since multi-label learning is a special case of multi-task learning \cite{read2023multi, huang2013multi}, the gradient directions of different labels may conflict, making it difficult for the model to optimize the performance of all labels simultaneously. An effective solution to these problems is to explore the correlations between labels \cite{Zhang_Zhou_2014, Liu_Wang_Shen_Tsang_2022, yeh2017learning, Zhang_Zhang_2010}. The potential correlations among labels are intricate, which can be pairwise, involve triples of labels, or even be common across all labels. Prompt learning is an emerging paradigm in deep learning, offering significant flexibility in task adaptation \cite{LiuPL2023}. Recent studies \cite{gppt, AllinOne, GraphPrompt, GraphPromptTuning} have introduced prompt learning into the field of GNNs. These methods generally employ soft prompts to learn task information, revealing the potential of prompt learning in enhancing task information capture capabilities of GNNs. Therefore, using a graph prompt method to model complex task correlations appears beneficial.

\begin{figure}[t]
\centering
\includegraphics[scale=0.4]{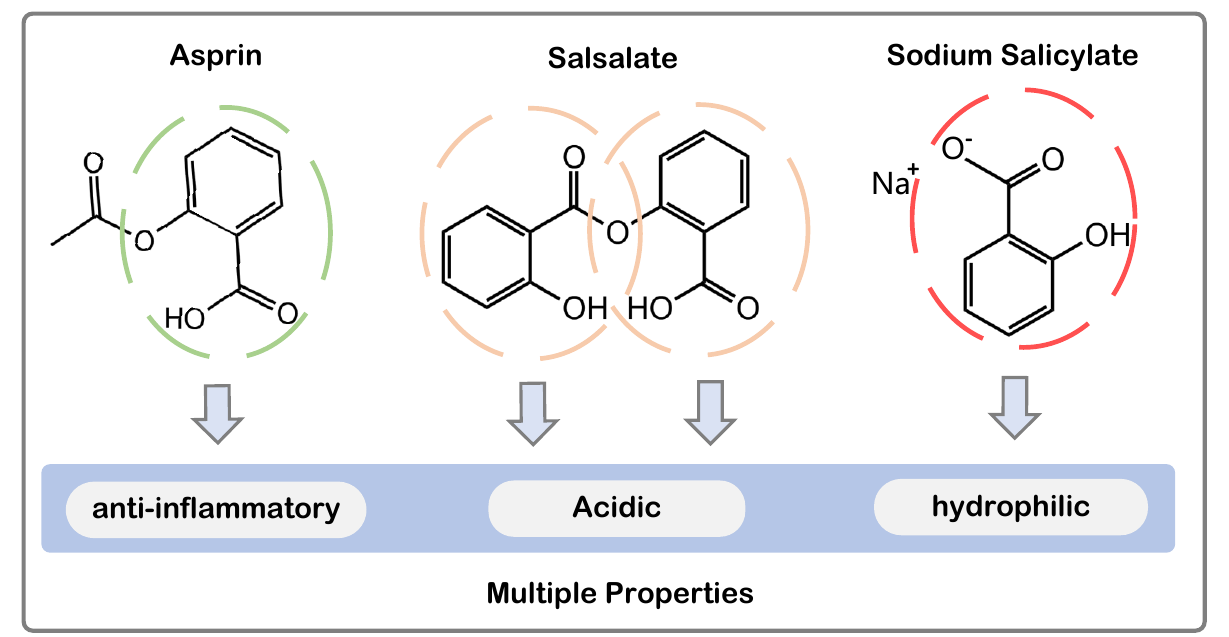}          
\caption{The salicylic acid structures in aspirin, salsalate, and sodium salicylate are specially emphasized by dashed circles, which collectively imply that these compounds are anti-inflammatory, acidic, and hydrophilic.}
\vspace{-20px}
\label{fig1} 
\end{figure}

Furthermore, another issue specific to molecular property prediction tasks is that \pagebreak certain motifs \cite{milo2002network} cause molecules to exhibit multiple properties in many cases. For example, as shown in Fig. \ref{fig1}, the salicylic acid structure is a common motif in Non-steroidal Anti-inflammatory Drugs (NSAIDs), indicating that drugs containing this structure have anti-inflammatory property. Meanwhile, the carboxyl group (-COOH) in the salicylic acid structure imparts hydrophilicity and acidity to these compounds. Thus, motif information is critical for multi-label molecular property prediction tasks. 

In this paper, we propose HiPM, an innovative framework to address the challenge of multi-label molecular property prediction. HiPM comprises two modules: the Molecular Representation Encoder (MRE) and the Task-Aware Prompter (TAP). MRE employs a hierarchical network architecture to learn molecular chemical structure features at both the atom and motif levels. TAP utilizes a hierarchical prompt tree to model multi-granular task correlations, employing learnable soft prompts and agglomerative hierarchical clustering to construct the tree. Ultimately, HiPM generates molecular representations that encapsulate differential task correlation information.

We conducted performance comparison experiments on six multi-label datasets from MoleculeNet \cite{wu2018moleculenet}, and the results indicate that our method achieves state-of-the-art performance across all datasets, attaining the best results on five of them. Additionally, we provide extensive supplementary experiments to offer a deeper understanding of our method.

In conclusion, our contributions can be summarized as follows:
\begin{itemize}
\item To the best of our knowledge, HiPM is the first to apply a prompt-based method to model multi-label task correlations in the field of molecular representation learning.
\item We designed the Task-Aware Prompter (TAP), which enables the model to adaptively learn multi-granular task correlation information.
\item We conducted comprehensive experiments to demonstrate that our framework achieves state-of-the-art performance in multi-label molecular property prediction scenarios and exhibits excellent interpretability.
\end{itemize}

\section{Related Work}\label{sec2}
Many graph-based methods have been proposed for molecular property prediction. For instance, Gilmer et al. \cite{MPNN2017} introduced the message passing neural networks (MPNNs) framework, which unifies existing models for graph data and enhances molecular property predictions. Wang et al. \cite{Automated3D} developed a pre-training framework for 3D molecular graphs to obtain comprehensive representations. Lv et al. \cite{Meta-GAT, Lv2022Meta-MolNet} leveraged meta-learning with graph attention networks to capture local atomic group effects and their interactions. Jiang et al. \cite{PharmHGT} proposed a pharmacological constraint-based multi-view molecular representation graph to extract significant chemical information from functional substructures and reactions. Several prompt-based methods have also been introduced into the field of molecular property prediction. Guo et al. \cite{Guo2024MolTailorTC} used the language model as an agent to highlight task-relevant features by understanding natural language descriptions. Fang et al. \cite{fang2023knowledge} developed a molecular contrastive learning framework using functional prompts from a knowledge graph. Despite these advancements, they fall short in effectively addressing the challenges of multi-label molecular property prediction, without considering the correlations between tasks. In contrast, our method uses prompt learning to capture task correlations in multi-label scenarios. To the best of our knowledge, our work is the first to apply prompt learning to model multi-label task correlations in molecular representation learning.

\section{Method}\label{sec3}
\subsection{Overview of HiPM}
\begin{figure*}[htp]
    \centering
    \includegraphics[width=1.01\linewidth]{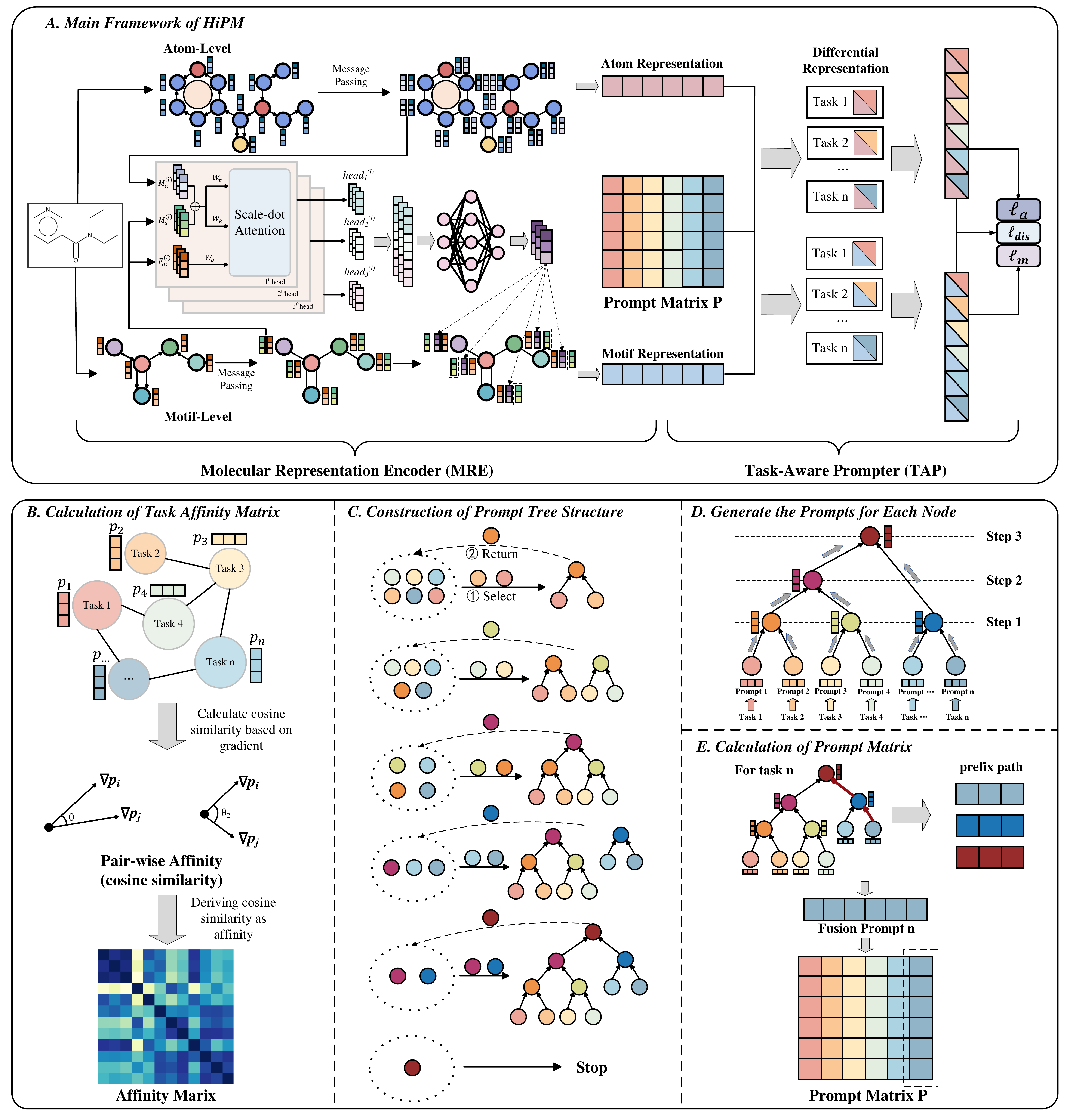}
    \caption{Overview of the HiPM Framework: (A) illustrates the architecture of our model, where the prompt matrix is directly fused with the molecular representations generated using Eq \ref{eqIns2}. (B) details the process of calculating task affinities using cosine similarities. (C) describes the methodology for constructing the hierarchical prompt tree structure, utilizing the agglomerative hierarchical clustering algorithm. (D) outlines the process of computing soft prompts for non-leaf nodes in a bottom-up manner. (E) explains how the prompt matrix is derived from the nodes corresponding to the prefix paths of tasks on the prompt tree. We use Eq. \ref{eqPro} to perform fusion on all prompts along the path.}
    \label{fig_Overview}
\end{figure*}

In this section, we provide an overview of HiPM. As illustrated in Fig. \ref{fig_Overview}, our framework consists of two primary components: the Molecular Representation Encoder (MRE) and the Task-Aware Prompter (TAP).

MRE is a hierarchical architecture that incorporates the GNN layer for each level and an information interaction module between different levels. Specifically, MRE captures molecular information using Message Passing Neural Network (MPNN) \cite{MPNN2017} at both the atom and motif levels. Following each message passing phase, MRE employs a Transformer-based Local Augmentation module \cite{himgnn} to integrate information from these two levels and generate the augmented motif-level message. Ultimately, MRE produces a molecular representation that encapsulates molecular chemical and structural information from both levels through a readout process.(Fig. \ref{fig_Overview}A)

TAP is designed to facilitate the joint learning of shared information across tasks at multiple granularities. In this module, each task is initialized with a learnable soft prompt, and their affinity is measured using cosine similarity. By calculating the affinity between all soft prompts, an affinity matrix is obtained (Fig. \ref{fig_Overview}B). Under the guidance of task affinity, TAP applies agglomerative hierarchical clustering algorithm to construct a tree structure (Fig. \ref{fig_Overview}C). Each leaf node in the tree corresponds to a specific task. Subsequently, TAP calculates the prompt for each node in a bottom-up manner (Fig. \ref{fig_Overview}D). Finally, TAP constructs the prompt matrix $\tilde{P}$ using the prompts along the path from the root node to each leaf node (Fig. \ref{fig_Overview}E).

In summary, HiPM integrates the prompt matrix $\tilde{P}$ with the molecular representations module to generate molecular representations that encapsulate task correlations.

\subsection{Molecular Representation Encoder (MRE)}

The hierarchical message passing mechanism allows MRE to generate effective molecular representations, which have been extensively proved to be simple and effective for learning multi-level molecular structural information \cite{himgnn,himol,pmlr-v119-jin20a,mgssl}. The following sections provide the details of MRE.

\subsubsection{MPNNs for Atom and Motif Levels}

In both the atom and motif levels, a molecule can be abstracted as a graph $G = (V, E, F_n, F_e)$, where $|V| = N_n$ denotes a set of $N_n$ nodes, $|E| = N_e$ denotes a set of $N_e$ edges, $F_n \in \mathbb{R}^{N_n \times d_n}$ denotes the node features, and $F_e \in \mathbb{R}^{N_e \times d_e}$ denotes the edge features. Here, $ d_n $ represents the dimension of the node features, and $ d_e $ represents the dimension of the edge features. At the atom level, atoms are represented as nodes and chemical bonds as edges. At the motif level, motifs are treated as nodes, and the overlapping atoms between motifs are treated as edges. Molecular motifs are extracted using an established algorithm from the literature \cite{ji2022relmole}. Each level utilizes a one-layer MPNN \cite{MPNN2017} to learn molecular features. Through the message passing process, MRE captures information at each level, enabling the fine-grained learning of molecular features.

\subsubsection{Information Interaction between Different Levels}
Learning molecular representations at isolated levels can result in the loss of local information. To improve the interaction between motifs and atoms, MRE employs a Transformer-based Local Augmentation module, which facilitates the interaction of information across different levels.

Let $M_a^{(l)}$ denote the messages from the atom level generated by the $l$-th step of message passing, $M_s^{(l)}$ denote the messages from the motif level and $F_m$ denote the features of motif nodes. Here, $M_a \in \mathbb{R}^{N_a \times d_a}, M_s \in \mathbb{R}^{N_m \times d_m}, F_m \in \mathbb{R}^{N_m \times d_f}$ and $N_a$ denotes the atom number, $N_m$ denotes the motif number, $d_a$ and $d_m$ denote the dimension of the message from the atom and motif levels, while $d_f$ denotes the dimension of the feature of motif. 

First, in the style of the Transformer \cite{attn}, three separate linear layers are employed to learn $Q^{(l)}$, $K^{(l)}$, and $V^{(l)}$ respectively, which can be formulated as:

\begin{equation}
\left\{
\begin{aligned}
Q^{(l)} &= F_m^{(l)}W_q\\
K^{(l)} &= (M_a^{(l)} \oplus M_s^{(l)})W_k\\
V^{(l)} &= (M_a^{(l)} \oplus M_s^{(l)})W_v
\end{aligned}
\right.
\end{equation}
where $\oplus$ denotes the concatenation, $W_q$, $W_k$, and $W_v$ are the projection matrix for $Q$, $K$, and $V$, respectively. 

Subsequently, the multi-head attention mechanism is employed to facilitate information interaction across the levels, and the corresponding formulas are as follows:
\begin{equation}
head_i^{(l)}=softmax(\frac{Q^{(l)}{K^{(l)}}^\top}{\sqrt{d_k}})V^{(l)}
\end{equation}
\begin{equation}
M_m^{(l+1)}=(head_1^{(l)} \oplus head_2^{(l)} \oplus...\oplus head_n^{(l)})W_o
\end{equation}
where $head_i$ denotes the attention computed by the $i$-th head, $d_k$ denotes the dimension of features that the $k$-th head processes, and $W_o$ is the projection matrix of multi-head feature concatenation to generate the new motif message $M_m^{(l+1)}$ with atom-level and motif level information.

\subsection{Task-Aware Prompter (TAP)}
TAP is designed to generate prompt matrix that containing multi-granular task correlation information. At the end of each training epoch, it constructs a hierarchical prompt tree. Based on the hierarchical tree, a prompt matrix is built for the forward propagation in next training epoch. Further details are provided in the following sections.

\subsubsection{Measure of Inter-Task Affinity}
To effectively reflect the conflict situations during overall training, we quantify task affinity using the cosine similarity of the cumulative prompt gradients over one training epoch. 

Specifically, we use an affinity matrix $A$ to describe the affinity among tasks, where $A_{ij}$ denotes the affinity between the $i$-th and $j$-th tasks. The soft prompt corresponding to the $i$-th task is defined as $p_i$, where $p_i \in \mathbb{R}^{d_r}$, with $d_r$ denoting the dimension of the molecular representation from MRE. Each soft prompt is initialized with a normal distribution, $p_i \sim \mathcal{N}(\mu, \sigma^2 I)$, where $\mu$ is the mean vector and $\sigma^2 I$ is the covariance matrix. For the $i$-th and $j$-th tasks, their affinity is calculated by Eq. \ref{cos_sim}:
\begin{equation} \label{cos_sim}
A_{ij}=\frac{{\nabla p_i}^\top \nabla p_j}{\|\nabla p_i\| \|\nabla p_j\|}
\end{equation}
where $\nabla p_i$ represents the cumulative gradient of the soft prompt for the $i$-th task, and $\|\nabla p_i\|$ denotes the $\text{L}_2$ norm of $\nabla p_i$.

\subsubsection{Construction of Hierarchical Prompt Tree}
The construction of hierarchical prompt tree is decomposed into two steps: constructing the structure of the prompt tree and generating the prompts for each node.

Agglomerative hierarchical clustering algorithm naturally forms a binary tree structure, providing an effective method for constructing prompt trees. Since the clustering process is typically based on distance, we need to transform the affinity matrix (cosine similarity matrix) $A$ to a cosine distance matrix $\tilde{A}$. The transformation method is as follows: 

\begin{equation}
    \tilde{A}=1-A 
\end{equation}

Then, the structure of the prompt tree is obtained through the following process. Considering $m$ tasks with the corresponding cumulative prompt gradients $\nabla p_1, \nabla p_2, \ldots, \nabla p_m$ obtained over one training epoch, initially, each gradient vector $\nabla p$ is treated as an individual cluster $c$, forming a set $\mathcal{C}$. In each iteration, the algorithm identifies the two closest clusters in $\mathcal{C}$, merges them into a new cluster, and updates $\mathcal{C}$ accordingly. Specifically, if the clusters $c_i$ and $c_j$ are merged into $c_k$, then $c_i$ and $c_j$ are removed from $\mathcal{C}$, and $c_k$ is added. This merging process continues until only one cluster remains in $\mathcal{C}$, at which point the algorithm terminates.

It is important to note that $\tilde{A}$ only provides the distances between initial clusters. We employ Eq. \ref{eqClusterd} to calculate the distance between clusters not included in $\tilde{A}$. In this equation, $d(u,v)$ denotes the distance between two clusters, $u$ and $v$. $u_i$ and $v_j$ refer to the points in clusters $u$ and $v$. Additionally, $|u|$ and $|v|$ denote the number of points of clusters $u$ and $v$, respectively.

\begin{equation}\label{eqClusterd}
    d(u,v)=\sum_{{u_i \in u, v_j\in v}}\frac{d(u_i,v_j)}{|u||v|}
\end{equation}

Based on the tree structure, we adopted a recursive method for calculating the prompts for tree nodes. Let the prompt tree be denoted as $T$. First, for the leaf nodes $T_i$ of the binary tree, we assign a learnable vector as its corresponding soft prompt $p_i$, where $1 \le i \le m$, $m$ is the total number of tasks. In the prompt tree, a non-leaf node $T_f$ must have a left child node $T_l$ and a right child node $T_r$ due to the nature of agglomerative hierarchical clustering. Then, the soft prompt of $T_f$ can be calculated by Eq. \ref{agg}, where $p_f$, $p_l$, and $p_r$ are the soft prompts of the $T_f$, $T_l$, and $T_r$, respectively. 
\begin{equation} \label{agg}
    p_f = p_l + p_r
\end{equation}

Through this recursive structure, we can calculate the soft prompt corresponding to each node of the tree from bottom to the top. For leaf nodes, the soft prompts contain individual information related to their corresponding tasks. For non-leaf nodes, the soft prompts contain shared information from their descendant nodes. The higher a node is positioned in the prompt tree, the more labels it encompasses. Therefore, the granularity of task information contained in the soft prompts of each tree node varies.

\begin{algorithm}
    \caption{Construction of Prompt Matrix $\tilde{P}$}\label{algo1}
    \textbf{Input:} Gradient matrix $G$, distance matrix $\tilde{A}$, mapping $\mathcal{M}$ of leaf nodes to their prefix paths \\
    \textbf{Output:} Prompt matrix $\tilde{P}$
    \begin{algorithmic}[1] 
        \Function{FindPrefix}{$k$, $path$, $\mathcal{M}$}
            \State $path \gets path \cup \{k\}$
            \If {$\mathcal{M}[k] = \emptyset$}
                \State $\mathcal{M}[k] \gets path$
            \Else
                \State \Call{FindPrefix}{$k_l$, $path$, $\mathcal{M}$}
                \State \Call{FindPrefix}{$k_r$, $path$, $\mathcal{M}$}
            \EndIf
        \EndFunction
        \\
        \Function{ConstructP}{$G$, $\tilde{A}$, $\mathcal{M}$}
            \State $\mathcal{L} \gets \Call{HierarchicalCluster}{G, \tilde{A}}$
            \State $P_0 \gets \emptyset$
            \ForAll{$(l, r) \in \mathcal{L}$}
                \State $p_f \gets p_l + p_r$
                \State $P_0 \gets P_0 \cup \{p_f\}$
            \EndFor
            \State \Call{FindPrefix}{$root$, $\emptyset$, $\mathcal{M}$}
            \State $P \gets \emptyset$
            \ForAll{$k \text{ in } 1,2,3,...,m$}
                \State $\mathcal{S} \gets \mathcal{M}[k]$
                \State $p_k \gets \sum_{p_i \in P_0}{\mathbb{I}(p_i \in \mathcal{S})p_i} / |\mathcal{S}|$
                \State $P \gets P \cup \{p_k\}$
            \EndFor
            \State // Assemble $P$ into a matrix $\tilde{P}$
            \State \Return $\tilde{P}$
        \EndFunction
    \end{algorithmic}
\end{algorithm}

\subsubsection{Calculation of Prompt Matrix}
During model inference, it is essential to consider the information of a specific task at all levels of granularity. To address this problem, we generate a new soft prompt $p_i'$ containing multi-granular task information for each task.

For any task $t_k$, with its corresponding leaf node $T_k$, the prefix path of $T_k$ is defined as the set consisting of all nodes unidirectionally connected to it, which is denoted as $pre$. We use the \textproc{FindPrefix} in Algorithm \ref{algo1} to compute prefix path of each task. Then, the new soft prompt for task $t_k$ is calculated by Eq. \ref{eqPro}, where $S = \{T_k\} \cup pre$, a node set including both the leaf node $T_k$ and all nodes in $pre$.

\begin{equation}\label{eqPro}
    p_k'=\frac{\sum_{j \in S}p_j}{|S|}
\end{equation}

For higher computational parallelism, we transform the set of soft prompts corresponding to all tasks \(P=\{p_1',p_2',p_3',...,p_m'\}\) into a prompt matrix \(\tilde{P}\) for forward propagation, where the \(i\)-th column of \(\tilde{P}\) represents the prompt for the \(i\)-th label, and \(m\) denotes the total number of labels. The process for calculating the prompt matrix $\tilde{P}$ is outlined in Algorithm \ref{algo1} and more details can be found in supplementary materials.

\subsubsection{Integration of Prompts and Molecular Representation}
We fuse the molecular representation produced by MRE with the prompts to generate the molecular representation including differential information across multiple granularities. Given the molecular representation \(r \in \mathbb{R}^{d_r}\) and the prompt matrix \(\tilde{P} \in \mathbb{R}^{d_r \times m}\), we use the Eq. \ref{eqIns1} and Eq. \ref{eqIns2} to integrate the multi-label prompt information into the molecular representation.
\begin{equation}\label{eqIns1}
    H=r\tilde{P}
\end{equation}
\begin{equation}\label{eqIns2}
    r^{'}=\sigma_2(\sigma_1(HW_1+b_1)W_2+b_2)
\end{equation}

Here, \(r^{'} \in \mathbb{R}^m\) represents the new molecular representation fused with multi-label prompt information. \(\sigma_1\) and \(\sigma_2\) are non-linear activation functions. \(W_1 \in \mathbb{R}^{m \times m}\), \(b_1 \in \mathbb{R}^{m}\), \(W_2 \in \mathbb{R}^{m \times m}\), and \(b_2 \in \mathbb{R}^{m}\) are trainable parameters. Since the \(i\)-th column of \(\tilde{P}\) represents the prompt corresponding to the \(i\)-th label, the \(i\)-th element of \(r^{'}\) is related only to the molecular representation \(r\) and the prompt corresponding to the \(i\)-th label. This method effectively integrates task correlation prompt information into the molecular representation, enhancing the model's capability to handle multi-label tasks. Finally, $r^{'}$ is directly fed into a classifier or regressor for prediction.

\subsubsection{Updating the Hierarchical Prompt Tree}
Updating the hierarchical prompt tree is essentially a process of reconstruction, which requires careful consideration of both the basis and the timing of updates.

\textit{Impact of gradient descent on the optimization of the tree structure.} Agglomerative hierarchical clustering operates on the principle of maximum similarity. We hypothesize that tasks with higher similarity are more likely to be grouped into the same cluster as model performance improves. Thus, our method for updating the prompt tree involves rerunning Algorithm \ref{algo1} based on the new task affinity and prompt gradients.

\textit{Timing for updating the prompt tree.} We reconstruct the prompt tree at the end of each epoch. This update frequency helps to keep the additional time overhead from clustering within an acceptable range and does not increase the instability of model performance. Through periodic reconstruction, the hierarchical prompt tree can be dynamically adjusted as the model learns, thereby better capturing the correlations between labels.

\subsection{Prediction and Loss}
In this section, we provide relevant details of the prediction and loss function. For the $i$-th label, HiPM yields predicted outcomes at different levels, $\hat{y}_{i,a}$ and $\hat{y}_{i,m}$, where $\hat{y}_{i,a}$ is from the atom level and $\hat{y}_{i,m}$ is from the motif level. To effectively align the feature spaces for different levels, we introduce an additional contrastive loss as described in Eq. \ref{ct}. Then, the loss for the $i$-th label is calculated using Eq. \ref{ls}.

\begin{equation}\label{ct}
    \text{c}_{i}=\frac{1}{N}\sum{(\hat{y}_{i,a}-\hat{y}_{i,m})^2}
\end{equation}

\begin{equation}\label{ls}
l_i = l_{i,a} + l_{i,m} + \lambda \cdot c_{i}
\end{equation}

Here, $l_{i,a}$ and $l_{i,m}$ represent the losses computed at the atom and motif level, respectively. $\lambda$ denotes the penalty strength of the contrastive loss. For classification tasks, we use the Binary Cross-Entropy (BCE) loss function, while for regression tasks, we use the Mean Squared Error (MSE) loss function.

To address label imbalance in classification, we weight the loss for each label. Specifically, for the $i$-th label, let $N$ denote the total number of molecules and $N_i$ denote the number of molecules possessing the $i$-th property label. The weight $w_i$ is then calculated as $w_i = \frac{N_i}{N}$. The final weighted loss is given by Eq. \ref{w_l}.

\begin{equation}\label{w_l}
    \mathbb{L}(\cdot) = \sum_{i=1}^{m} w_i \cdot l_i
\end{equation}

\section{Experiment}\label{sec8}
In this section, we evaluate the performance of HiPM across various multi-label datasets. Specifically, we aim to answer the following questions:

\begin{enumerate}

\item How does HiPM compare to state-of-the-art models on the multi-label molecular property prediction tasks?

\item How does the model performance differ when considering task correlation versus not considering it?

\item What are the task correlations obtained through clustering by TAP?

\item How does TAP affect the attention weights of key molecular substructures?

\item How does TAP affect the predicted probabilities of HiPM for each task?

\end{enumerate}

\subsection{Experimental Setup}\label{subsec1}
\subsubsection{Benchmark Datasets}\label{subsubsec1}
To evaluate the effectiveness of HiPM for multi-label property prediction, we conducted experiments on six multi-label datasets from MoleculeNet \cite{wu2018moleculenet}. These datasets include four classification datasets and two regression datasets, covering two categories: physiology and quantum mechanics. The statistics of these datasets are shown in Table \ref{table2}. For a molecule represented by a SMILES \cite{smiles} string, we convert it into a 2D topological graph using RDKit\footnote{\href{https://www.rdkit.org/}{https://www.rdkit.org/}} and DGL-LifeSci\footnote{\href{https://lifesci.dgl.ai/}{https://lifesci.dgl.ai/}} for further processing.

\begin{table*}[t]
\begin{minipage}[t]{\linewidth}
\caption{Performance comparison on six multi-label datasets under scaffold splitting. Each model is run with three random seeds, and we report the average ROC-AUC (classification) or MAE (regression) scores, along with the corresponding standard deviations.\label{table1}}
\renewcommand{\arraystretch}{1.1}
\centering
\setlength{\tabcolsep}{0.02\textwidth}{
\resizebox{1.0\linewidth}{!}{
\scriptsize
\begin{tabular}{l|cccc|cc}
\hline
Type & \multicolumn{4}{c|}{Classification(ROC-AUC)} & \multicolumn{2}{c}{Regression(MAE)} \\
\hline
Dataset & Clintox($\uparrow$) & SIDER($\uparrow$) & Tox21($\uparrow$) & ToxCast($\uparrow$) & QM8($\downarrow$) & QM9($\downarrow$) \\
\hline

MPNN  & $0.879\pm0.054$ & $0.595\pm0.030$ & $0.808\pm0.024$ & $0.691\pm0.013$ & $0.0146\pm0.0015$ & $5.253\pm0.465$\\

D-MPNN  &$0.879\pm0.040$ & $0.610\pm0.027$ & $0.808\pm0.023$  & $0.718\pm0.011$ & $0.0124\pm0.0013$ & $4.797\pm0.343$\\
ML-MPNN  &$0.865\pm0.027$ & $0.609\pm0.017$ & $0.796\pm0.021$  & $0.708\pm0.003$ & $0.0225\pm0.0005$ & $5.734\pm0.581$\\

FP-GNN  &$0.765\pm0.038$ & $0.598\pm0.014$ & $0.803\pm0.024$  & $0.694\pm0.014$ & $0.0165\pm0.0019$ & $6.491\pm0.367$\\
MVGNN  &$0.894\pm0.035$ & $0.647\pm0.022$ & \underline{$0.835\pm0.010$}  & $0.736\pm0.011$ & \underline{$0.0125\pm0.0001$} & \textbf{2.347$\pm$0.018}\\
HimGNN & $0.922\pm0.037$ & $0.623\pm0.033$ & $0.803\pm0.028$  & $0.732\pm0.021$ & $0.0211\pm0.0065$ & $6.434\pm0.908$\\
\hline
GROVER  & $0.734 \pm 0.032$ & $0.567 \pm 0.031$  & $0.736 \pm 0.022$  & $0.593 \pm 0.021$ & $0.0184 \pm 0.0009$ & $6.497 \pm 1.356$\\
MolCLR  &$0.900\pm0.025$ & $0.597\pm0.009$ & $0.811\pm0.026$  & $0.691\pm0.012$ & $0.0182\pm0.0001$ & $3.215\pm0.106$\\
MGSSL  & $0.877\pm0.021$ & $0.591\pm0.016$  & $0.788\pm0.022$  & $0.573\pm0.004$ & $0.0322\pm0.0033$ & $/$\\
HiMol  & $0.702\pm0.092$ & $0.593\pm0.022$ & $0.815\pm0.010$  & $0.692\pm0.014$ & $0.4232\pm0.0033$ & $3.355\pm1.229$\\
KANO  & $0.903\pm0.039$ & $0.593\pm0.026$ & $0.797\pm0.012$ & $0.696\pm0.020$ & $0.0148\pm0.0025$ & \underline{$2.544\pm0.257$}\\
\hline
HiPM (wo/pro) & \underline{$0.948\pm0.041$} & \underline{$0.666\pm0.006$} & $0.770\pm0.005$ & $0.735\pm0.006$ & $0.0171\pm0.0003$ & $5.344\pm0.112$\\
HiPM (wo/cls) & \textbf{0.969$\pm$0.009} & $0.653\pm0.007$ & $0.799\pm0.007$ & \underline{$0.751\pm0.006$} & $0.0336\pm0.0004$ & $8.352\pm0.476$\\
HiPM (Ours) & $0.928\pm0.014$ & \textbf{0.672$\pm$0.010} & \textbf{0.843$\pm$0.007} & \textbf{0.786$\pm$0.004} & \textbf{0.0117$\pm$0.0001} & $5.238\pm0.199$\\
\hline
\end{tabular}
}}
\end{minipage}
\begin{tablenotes}%
\item Note: $\uparrow$ means that the higher result is better and $\downarrow$ means that the lower result is better. Note that the MGSSL model on QM9 is too time consuming to finish in time, and its results are not presented. The best score in each column is in \textbf{bold} and the second best score is \underline{underlined}. 'HiPM(wo/pro)' and 'HiPM(wo/cls)' are variants of HiPM, as detailed in the \textit{Ablation Study} section. 
\end{tablenotes}
\end{table*}

\begin{table}[t]
\centering
\caption{Statistics of datasets.\label{table2}}
\renewcommand{\arraystretch}{1.0}
\begin{tabularx}{\linewidth}{>{\centering\arraybackslash}X>{\centering\arraybackslash}X>{\centering\arraybackslash}X>{\centering\arraybackslash}X}
\hline
Dataset & Tasks & Task Type & Molecules\\
\hline
ClinTox & 2 & Classification & 1478 \\
SIDER & 27 & Classification & 1427 \\
Tox21 & 12 & Classification & 7831 \\
ToxCast & 617 & Classification & 8575 \\
QM8 & 12 & Regression & 21786 \\
QM9 & 12 & Regression & 133885 \\
\hline
\end{tabularx}
\end{table}

\subsubsection{Baseline Models}
We compared HiPM with existing state-of-the-art methods, categorized into supervised and self-supervised methods. The supervised methods include MPNN \cite{MPNN2017}, D-MPNN \cite{DMPNN}, ML-MPNN \cite{MLMPNN_2022}, FP-GNN \cite{fpgnn}, MVGNN \cite{mvgnn}, and HimGNN \cite{himgnn}. The self-supervised methods consist of GROVER \cite{grover}, MolCLR \cite{MolCLR}, MGSSL \cite{mgssl}, HiMol \cite{himol}, and KANO \cite{fang2023knowledge}. Detailed information about these methods is available in the supplementary materials.

\subsubsection{Implementation Details}
For all datasets, we employ scaffold splitting, which is considered superior to random splitting and helps to prevent information leakage \cite{LV202394}. The datasets are divided into training, validation, and test sets in an 8:1:1 ratio. We use the widely adopted Optuna \cite{optuna_2019} framework for hyper-parameter tuning, performing 20 trials per dataset to identify the optimal parameters based on validation set performance. Detailed hyper-parameter information is available in our \href{https://github.com/zhousongh/HiPM}{code repository}. Model training is executed on two NVIDIA GeForce 1080Ti GPUs, with 60 epochs for classification tasks and 200 epochs for regression tasks.

\subsection{Performance Comparison (RQ1)}\label{subsec4}
In this section, we compare the performance of HiPM against various baseline models across six multi-label datasets. From Table \ref{table1}, several key observations can be made:

(i) HiPM shows better performance in multi-label scenarios. In classification tasks, HiPM outperforms all baseline models across all datasets, achieving an average ROC-AUC improvement of 3.6$\%$. In regression tasks, HiPM shows an average improvement of 6.4$\%$ over other baseline models on QM8.

(ii) HiPM performs relatively average on QM9 compared to the state-of-the-art models. QM9, which comprises quantum mechanical calculations for a large number of small organic molecules, is characterized by its intricate data distribution. This complexity likely renders HiPM less optimal for such scenarios.

(iii) HiPM is particularly effective in scenarios where there are significant correlations between labels (e.g., Tox21), or when dealing with a larger number of labels with more complex correlations (e.g., ToxCast). This effectiveness highlights HiPM's capability to capture and leverage the intricate relationships between multiple labels.

\begin{figure*}[t]
    \centering
    \includegraphics[width=1.0\textwidth]{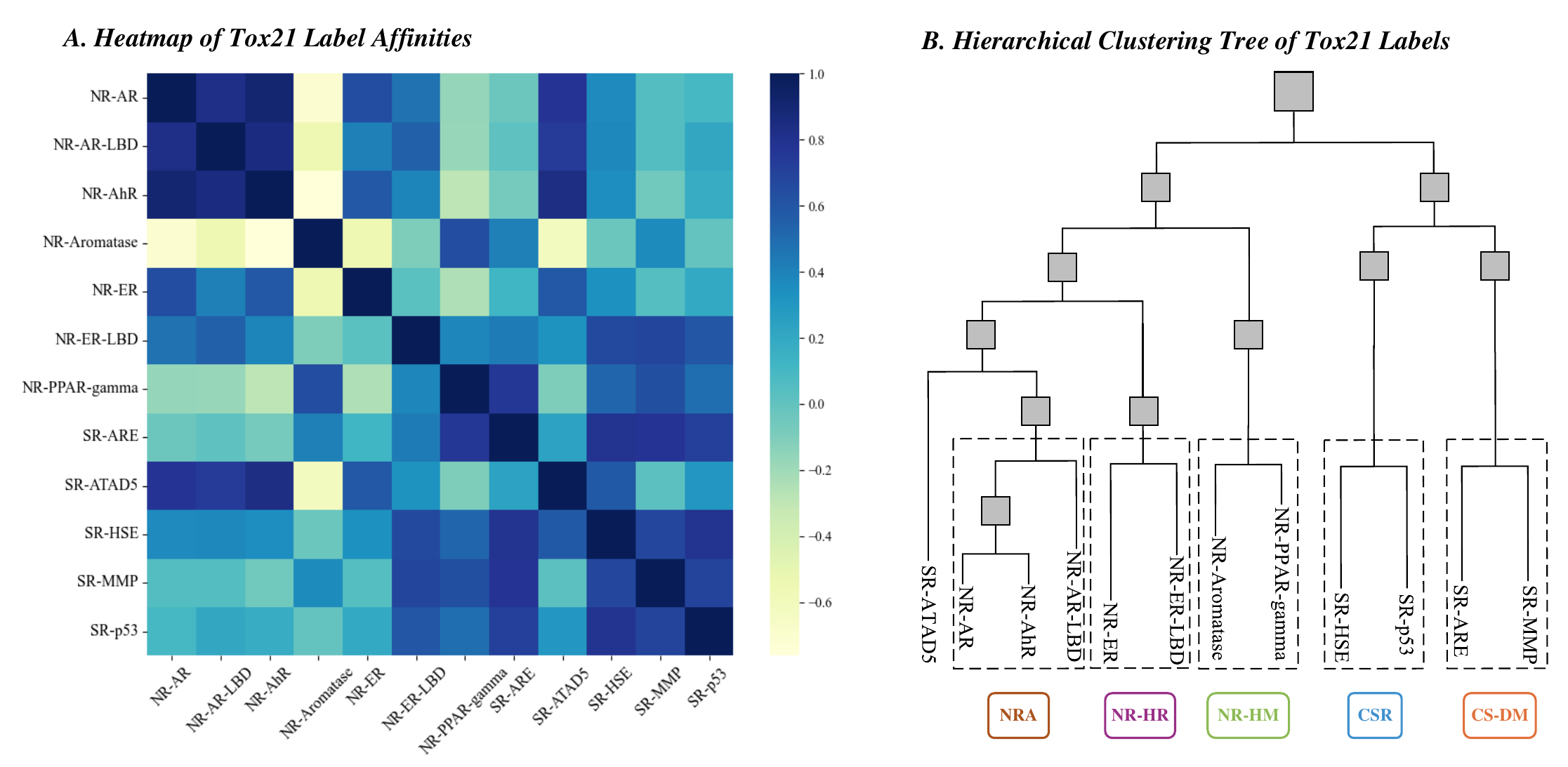}
    \caption{Visualization results of the affinity matrix (A) and hierarchical clustering tree (B) for the 12 tasks of Tox21. In the affinity matrix, darker colors represent higher task affinities. In the hierarchical clustering tree, the leaf nodes are labeled, and potential reasons for the prioritized clustering of certain tasks are provided. Specifically, CS-DM stands for Cellular Stress and Defense Mechanisms, CSR stands for Cellular Stress Response, NR-HM stands for Nuclear Receptor and Hormone Metabolism, NR-HR stands for Nuclear Receptor and Hormone Regulation, and NRA stands for Nuclear Receptor Activity. The clustering of these leaf nodes reflects their functional similarities and associations within their respective toxicological mechanisms.}
    \label{heatmap_and_tree}
\end{figure*}

\begin{figure}[t]
    \centering
    \includegraphics[scale=0.4]{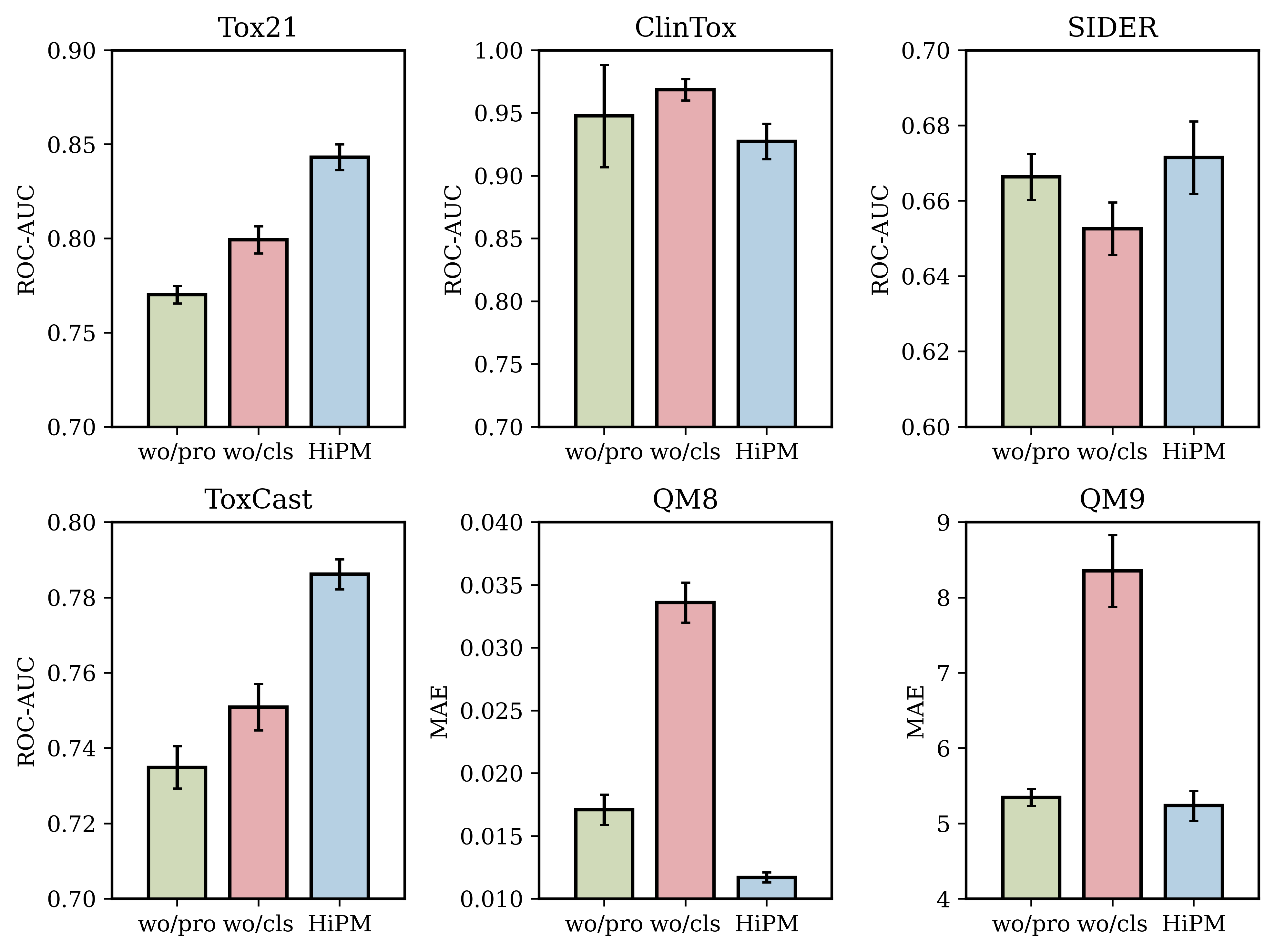}
    \caption{Results of ablation experiments. Each variant was run with three random seeds. We report the average ROC-AUC (classification) or MAE (regression) scores along with their standard deviations. Higher ROC-AUC values indicate superior performance, while lower MAE values are preferable. The error bars denote standard deviations.}
    \label{abla}
\end{figure}

\subsection{Ablation Study (RQ2)}\label{ablation_study}
In this section, we conducted ablation studies to further analyze the effectiveness of TAP. Specifically, we designed the following two variants of HiPM:

\begin{itemize}
    \item HiPM (wo/pro), which directly removes TAP. This variant reflects the effectiveness of MRE.
    \item HiPM (wo/cls), which includes TAP without clustering. This variant shows the model performance when only individual task information is considered.
\end{itemize}

For each dataset and variant, we used three random seeds for experiments, and the experimental results are shown in Table \ref{table1} and Fig. \ref{abla}.

\subsubsection{Comparison Between Considering and Not Considering Task Association}
The results show that HiPM achieves superior outcomes on five datasets compared to HiPM (wo/cls), reaching the optimal performance. The experimental results validate the significance of the clustering process in enhancing the model's ability to capture shared information. Furthermore, these results illustrate that multi-granular shared information contributes to improving model performance.

\subsubsection{Are Prompts Without Task Association Information Useful?}
The results reveal that HiPM (wo/cls) underperforms compared to HiPM (wo/pro) on half of the datasets. This suggests that focusing solely on individual task information may lead to conflicts. Therefore, excluding task information from the molecular representation entirely might be more beneficial when task correlations are not considered.

\begin{figure*}[t]
\centering
\includegraphics[width=1\linewidth]{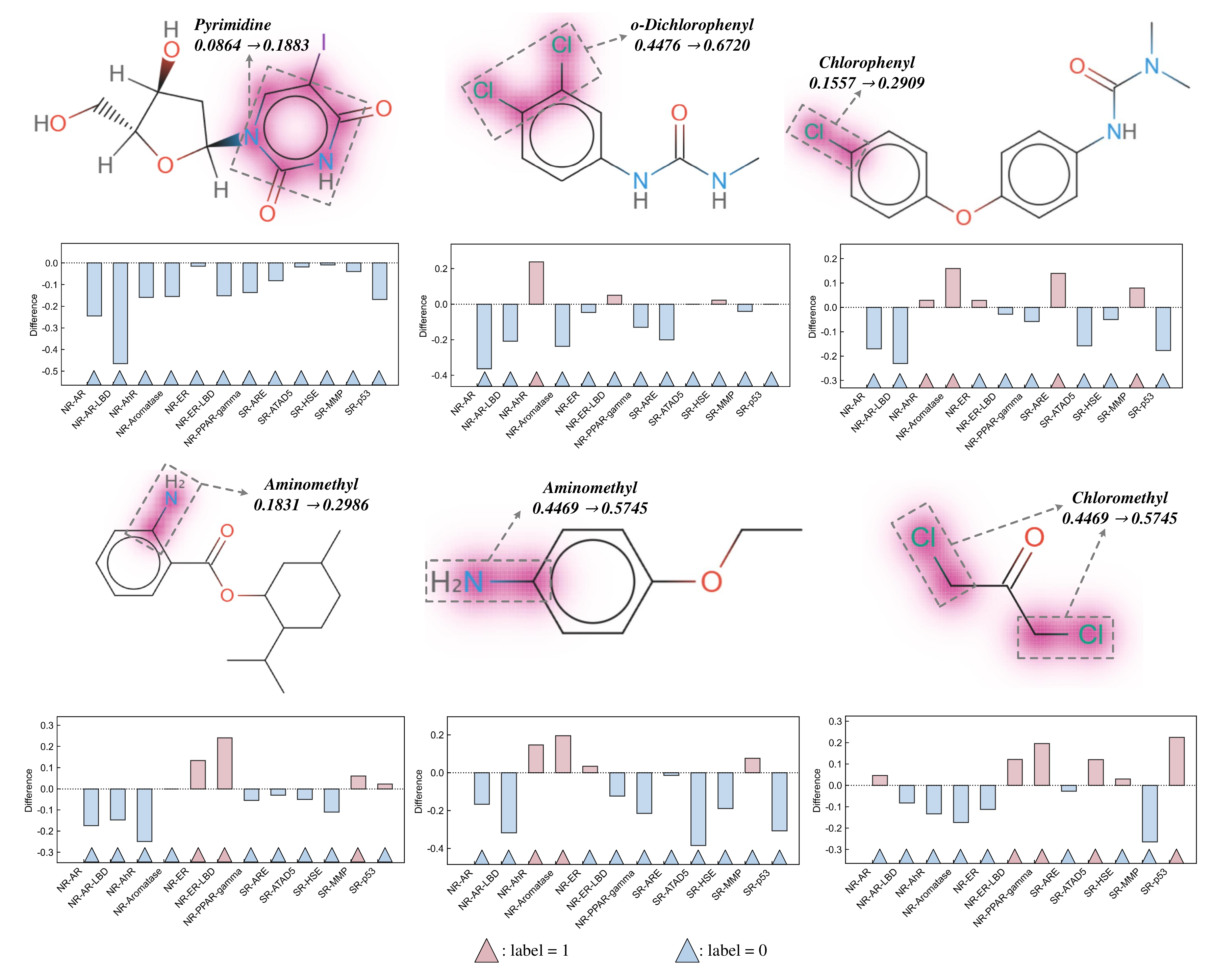}          
\caption{Visualization analysis results that highlight key motifs and the changes in the corresponding attention weights before and after integrating the TAP module. Bar charts illustrate the variations in predicted probabilities for each label generated by HiPM before and after the inclusion of TAP. The value of each bar is calculated as the predicted probability (after) minus the predicted probability (before). An increase (or decrease) in predicted probability after adding TAP, where the true label is 1 (or 0), indicates an increase in the model’s confidence in the correct answer.}
\label{mol_weight} 
\end{figure*}

\subsection{Interpretability Analysis (RQ3)}\label{subsec6}
In this section, we visualized the affinity matrix and hierarchical prompt tree for 12 tasks of Tox21. The labels in Tox21 are primarily categorized into Nuclear Receptor (NR) and Stress Response (SR) groups. Fig. \ref{heatmap_and_tree} demonstrates that dark colors are concentrated in the upper left and lower right corners of the matrix, indicating a high degree of task affinity within the NR and SR categories. For instance, the tasks NR-AR, NR-AR-LBD, and NR-AhR exhibit strong correlations and all belong to the NR category, demonstrating that our model can effectively capture the correlations between tasks. Conversely, the affinity between different categories is relatively low, such as between NR-Aromatase and SR-ATAD5, suggesting that our model can also capture the differences between tasks.

As shown in Fig. \ref{heatmap_and_tree}, the clustering results align closely with the affinity, prioritizing tasks with high affinity to be clustered into the same group. We have provided possible reasons for these clusters, which match known facts. This indicates that our agglomerative hierarchical clustering method possesses good interpretability. Additional information on other datasets can be found in the supplementary materials.


\subsection{Visualization Analysis (RQ4 \& RQ5)}\label{subsec7}

In this section, we selected several molecules from Tox21 to observe changes in molecular motif weights before and after incorporating the TAP module. Fig. \ref{mol_weight} illustrates the changes in molecular motif weights and predicted probabilities. For simplicity and clarity, only motifs with weight differences greater than $0.1$ were visualized.

Tox21 is a dataset focused on toxicity. After integrating the TAP module, the weights of certain toxicity-related motifs, such as aminomethyl and chlorophenyl, increased significantly, aligning the predicted probabilities more closely with the ground truths. This suggests that our model effectively captures key motif structures essential for these tasks.

\section{Conclusion}
In this paper, we present HiPM, a hierarchical prompted multi-label molecular representation learning framework developed to tackle the issues associated with multi-label molecular property prediction. HiPM leverages an innovative hierarchical prompt method, enabling the model to learn task-specific prompts and effectively capture correlation information across tasks. By constructing a hierarchical prompt tree based on task affinities, our method reveals latent multi-granular correlations among property labels. Overall, HiPM exhibits considerable potential in advancing multi-label molecular property prediction, offering a robust tool for drug discovery research. In future work, we aim to explore mechanisms for dynamically managing label associations across different datasets, which could enhance HiPM’s ability to adaptively generalize across a wide range of molecular property prediction tasks.


\subsection{Key Points}\label{subsec8}
\begin{itemize}
    \item We introduced HiPM, a hierarchical prompted molecular representation learning framework designed to address the complexity of multi-label molecular property prediction.
    \item HiPM employs a hierarchical prompt tree to model multi-granular task correlations, generating molecular representations that incorporate differential task information.
    \item By capturing multi-granular correlation information among labels, HiPM mitigates the negative transfer caused by conflicts of individual task information.
    \item Extensive experimental results demonstrate that HiPM exhibits strong competitiveness compared to existing state-of-the-art models.
\end{itemize}

\section{Data Availability}
Complete datasets and source code for HiPM are available for free on GitHub at \href{https://github.com/zhousongh/HiPM}{https://github.com/zhousongh/HiPM}.

\section{Author Contributions Statement}
L. K. and S. Z. contributed equally to this work. L. K. and S. Z. conceived the idea and initiated the project. L. K. and S. Z. designed and implemented the HiPM framework. S. L. supervised the project. L. K., S. Z., and S. F. tested the performance of the HiPM framework and compared it with state-of-the-art methods. L. K., S. Z., S. F., and S. L. drafted the manuscript. All authors reviewed and approved the final manuscript.

\section{Funding}
This work was supported by the National Natural Science Foundation of China (62102158); Huazhong Agricultural University Scientific \& Technological Self-innovation Foundation (BC2024108); Fundamental Research Funds for the Central Universities (2662021JC008, 2662022JC004). The funders have no role in study design, data collection, data analysis, data interpretation, or writing of the manuscript.

\bibliographystyle{unsrt}
\bibliography{paper}

\end{document}